# AERODYNAMIC MODELS FOR HURRICANES

## IV. On the hurricane genesis and maturing

A.I. Leonov

The University of Akron, Akron, OH 44325-0301

E-mail: **leonov@uakron.edu**

**Abstract**

This, fourth paper of the series (see previous papers in Refs.[1-3]) derives approximate equations for future numerical studies of initial evolution of hurricanes, develops a new analytical model of hurricane genesis and maturing, and presents simple results, which seem to be in accord with observations. Several remarks on tornados are also made at the end of the paper.

## 4.1. Introduction

The hurricane genesis, which includes the hurricane occurrence and maturing, is the most intriguing problem in studies of hurricanes. There are many observations of various hurricanes during period of their emergence and maturing (e.g. see [4], Ch.2.4.2 among others). There are also several papers [5-8] which proposed intrinsic mechanisms of hurricane occurrence and grow. The most important element of these models (see also [4]) are based on observations of existence of a threshold value of tangential wind speed. Only after overcoming this threshold the hurricane is maturing, i.e. growing in radial size with increasing wind speed. Unfortunately, the models [5-8] use turbulent approximations with unknown radial turbulent diffusivity coefficient, which is necessary to fit the observed values. For example model [7, 8] describes the final size of growing hurricane close to the observed values, only when the radial turbulent diffusivity of angular momentum is properly chosen. Thus in spite of the results in [5-8], the physical aspects of hurricane genesis and maturing still remain vague. Paper [9] mostly devoted experimental studies of the hurricane mesovortices, also discussed very important role of Kelvin-Helmholtz instability in the initiating and functioning hurricanes.

The essential feature of hurricanes is that they are initiated in the near-equatorial zones, but not on equator. It indicates that pre-conditions for their emergence are the presence of extremely hot air, and importance of Coriolis force. One could also notice that in summer-fall times specific winds exist in these regions, which have high enough level of horizontal gradients. Based on these observations, the following 3-step scenario of hurricane genesis is proposed.

- The *first step*: a *plume* of warm and humid air emerges and moves vertically upward.
- The *second step*: the emerged vertical plume captures the rotation from a horizontally sheared wind with a following restructuring of the emerged vertical jet. In this step, the plume acquires an initial value of full angular momentum $M_0$.
- The *third step*: the hurricane grows in the radial direction. This grow is caused by the Kelvin-Helmholtz instability with radial propagation into environmental air under action of Earth rotation, until the hurricane radius reaches the value at which the relative angular velocity completely decays.

Although consequential character of this scenario might be doubtful, it is still convenient to analyze the above three steps separately as simple consequence of these events.

The first step is caused by very warm, subtropical locations of emerging hurricanes, where many plumes could emerge. The second step provides in seldom occasions the initial vorticity for the emerged plumes. And the third step being consistent with the viewpoint of paper [9], involves a specific, however, linearized mechanism of radial propagation of Kelvin-Helmholtz instability, with Coriolis force playing important role.

The present paper develops a new prognostic mechanism for these phenomena based on the above scenario. The paper is organized as follows. Section 4.2 derives an approximate set of unsteady aerodynamic equations for initial stage of developing hurricane. It is based on the averaged "jet" approach used in paper [2] for stationary situation. The distinct feature of these equations is that they ignore existence of the eye region of hurricane at the initial stage its development. These equations are needed not only for future numerical investigations of hurricane restructuring but also for obtaining important scaling relations. Section 4.3 briefly analyzes emerging plume and its restructuring to initial premature hurricane. Section 4.4 proposes and develops a simple analytical model for hurricane propagation and maturing. The results of model evaluations seem to be quite realistic. Finally Section 4.5 of the paper discusses possible measures for hurricane suppression and applications of the developed models for description of tornados.

## 4.2. Initial stage of hurricane evolution

In order to describe this stage we have to formulate the equations for ascending and rotating warm jet emerged in the relatively cold air environment, and then consider as a non-rotating limit, the case of plume. As before, all the equations below describe axisymmetric distributions of dynamic variables using cylindrical coordinate system.

We consider now approximate dynamic equations for ascending and rotating jet, occupying the region $\{0 < r \leq r_e(z,t)\}$ and unrestricted in the vertical direction. The main physical assumption is that at initial stages of hurricane evolution, there is no eye region, which is developed later during a long period of hurricane maturing. Instead we assume the existence of quasi-rigid rotation of the core (or "jet") region of hurricane, $0 < r < r_e(z,t)$ with angular velocity independent of radius.

In the non-steady case, the unsteady extension of governing equations for the steady pseudo-adiabatic "jet approach", derived in [2], is presented as:

$$\partial_t(\rho_T s_e) + \partial_z(\rho_T s_e u) = 0, \quad s_e = r_e^2(z,t), \quad \rho_T \approx \rho_a T_a^0 / T_0. \tag{4.1}$$

$$\partial_t(\rho_T s_e u) + \partial_z(\rho_T s_e u^2) = \partial_z(p_a s_e - I_p) + g[(\rho_a - \rho_T)s_e], \quad I_p(z,t) \equiv 2\int_0^{r_e} rp\,dr. \tag{4.2}$$

$$\partial_t(\rho_T s_e^2 \hat{\omega}) + \partial_z(\rho_T s_e^2 \hat{\omega} u) = 0, \quad \hat{\omega} = \omega(z,t) + \Omega. \tag{4.3}$$

Here $\Omega \approx 5.5 \times 10^{-6}$ is the local angular velocity of Earth rotation, $\rho_T$ and $\rho_a(z)$ are the densities in the warm jet and in the atmospheric environment, respectively, $T_a^0$ and $T_0$ are the jet and environmental temperatures at $z = 0$, respectively, and the jet variables depending on vertical $z$ coordinate and time $t$ are: the upward directed vertical velocity $u$, and the relative angular velocity $\omega$, and the jet radius $r_e$.

Utilizing now (4.1), equations (4.2) and (4.3) can be rewritten as:

$$\partial_t u + u \partial_z u = \frac{1}{\rho_T s_e} \partial_z (p_a s_e - I_p / 2) + g \frac{T_0 - T_a^0}{T_a^0}, \quad I_p(z,t) \equiv 2 \int_0^{r_e} rp\,dr \quad (4.2a)$$

$$\partial_t M + u \partial_z M = 0, \quad M_e(z,t) = (\omega + \Omega) s_e. \quad (4.3a)$$

In equation (4.3a), $M_e(z,t)$ is the full angular momentum at the radius $r_e(z,t)$ of jet. In the present and next Sections we neglect the effect of $\Omega$, since even at the beginning of jet occurrence, the initial value $\omega_0$ is very large as compared to $\Omega$, i.e. $\omega_0 \gg \Omega$. In this case $M(r,z,t) = M_e(z,t) r^2 / r_e^2$ when $0 < r \leq r_e(z,t)$.

We now consider the exterior region of jet, $r_e(z,t) \leq r \leq r_a(t)$, which propagates in the radial direction with the speed $\dot{r}_a(t)$, assuming for simplicity that $r_a$ is $z$ independent. When neglecting the effects of both the radial and vertical velocities in this region, rotating jet induces rotational motion outside the jet. Simplifying model assumption is that the radial distribution of angular momentum in this region is similar to the quasi-stationary case, i.e. $M(r,z,t) \approx M_e(z,t)$ $(r > r_e)$. Thus the continuous radial distribution of angular momentum in the whole radial region is:

$$M(r,z,t) \approx M_e(z,t) \times \begin{cases} (r/r_e)^2, & 0 < r \leq r_e \\ 1, & r_e \leq r \leq r_a \end{cases}. \quad (4.4)$$

Neglecting also effects of radial and vertical velocities on pressure in the entire region reduces the radial momentum balance (1.8) from [1] to equation for pressure distribution:

$$\frac{\partial p}{\partial r} \approx \rho \frac{M^2}{r^3} = \rho_a M_e^2 \times \begin{cases} (T_0/T_a^0) \cdot (r/r_e^4), & 0 < r \leq r_e \\ 1/r^3, & r_e \leq r \leq r_a \end{cases}$$

Integrating this equation with continuity condition at $r = r_e$ and boundary condition, $p = p_a$ at $r = r_a$ yields:

$$p = p_a - \frac{\rho_a M_e^2}{2r_e^2}\left[1 + \frac{T_0}{T_a^0}\left(1 - \frac{r^2}{r_e^2}\right) - \frac{r^2}{r_a^2}\right] \quad (0 \le r \le r_e)$$

$$p = p_a - \frac{\rho_a M_e^2}{2r_e^2}\left(\frac{r_e^2}{r^2} - \frac{r_e^2}{r_a^2}\right) \quad (r_e \le r \le r_a) \quad (4.5)$$

Calculating $I_p$ defined in (4.2) yields:

$$I_p = p_a s_e - \rho_a \frac{M_e^2}{2}\left(1 - \frac{T_0}{2T_a^0} - \frac{s_e}{s_a}\right), \quad (s_a = r_a^2). \quad (4.6)$$

Using adiabatic formula $\rho_a = \rho_a^0 (1-\hat{z})^{2.5}$ where $\hat{z} = gz/(3.5RT_a^0)$ is barometric altitude, one can rewrite the governing equations (4.1), (4.2a) and (4.3a) in the final form:

$$\partial_t[(1-\hat{z})^{2.5} s_e] + \partial_z[(1-\hat{z})^{2.5} s_e u] = 0 \quad (4.7a)$$

$$\partial_t u + u \partial_z u = \frac{T_0/T_a^0}{2(1-\hat{z})^{2.5} s_e} \partial_z[(1-\hat{z})^{2.5} M_e^2 (1 + 0.5 T_a^0/T_0 - s_e/s_a)] + g(T_0 - T_a^0)/T_a^0 \quad (4.7b)$$

$$\partial_t M + u \partial_z M = 0. \quad (4.7c)$$

Equations (4.7a-c) are closed set of averaged jet equations relative to cross-sectional jet area $s_e$, vertical velocity $u$ and angular momentum $M_e$ at the jet radius $r_e = \sqrt{s_e}$ if the function $s_a(t)$ is known. This is a subject of study in Section 4.4 of this paper. These equations describe the initial evolution of hurricane. Equation (4.7.a) describes the mass balance, (4.7b) the balance of vertical momentum, and (4.7c) the balance of angular momentum in the vicinity of jet. This set will be used elsewhere for numerical analyses.

Scaling approach applied to the set (4.7a-c) is useful for both the numerical studies and evaluating characteristic scales for time and vertical dimension for fast restructuring of hurricane. Recall that scaling in the steady equations for hurricane has been introduced in paper [2] of the series. Extending this stationary scaling for non-steady case, we introduce the non-dimensional variables denoted below by tildes, along with the buoyancy parameter $\tau$ as:

$$\hat{M}_e = \frac{M_e/\sqrt{2s_0}}{\sqrt{RT_a^0}}, \quad \hat{u} = \frac{u}{\sqrt{RT_a^0}}, \quad \hat{s}_e = \frac{s_e}{s_0}, \quad \hat{t} = \frac{g}{3.5\sqrt{RT_a^0}}t, \quad \hat{z} = \frac{gz}{3.5RT_a^0}, \quad \tau = 3.5\frac{T_0 - T_a^0}{T_a^0} \quad (4.8)$$

Here $s_0$ is a characteristic value of jet cross-section, e.g. area of initial bottom spot. In non-dimensional variables (4.8), equations (4.7a-c) are of the same form with changing notations of variables by those with tildes, and additionally, in equation (4.7b) the buoyancy term in the right-hand side changes for $\tau$.

Characteristic values of scales for time $t_0$, vertical size $z_0$, and vertical speed of propagation of disturbances $v_0$ are obtained from (4.8) as:

$$t_0 = \frac{3.5\sqrt{RT_a^0}}{g}, \quad z_0 = \frac{3.5 RT_a^0}{g}, \quad v_0 \approx z_0/t_0 = \sqrt{RT_a^0}. \tag{4.9}$$

Using standard values of parameters in (4.9) yields: $t_0 \approx 100\,\text{sec}$, $z_0 \approx 30\,km$, and $v_0 \approx 300\,m/s$. These figures could be interpreted as follows: any restructuring in hurricane quickly propagates throughout the entire hurricane height in about 100sec. When $t \gg t_0$, the hurricane could be considered as steady or quasi-steady. In this case, all the time derivatives in equations (4.7a-c) can be ignored.

## 4.3. Emerging plume and capture of angular momentum

We define plumes as vertical non-rotating jets caused by buoyancy and inertia effects in air. Then in non-dimensional variables (4.8) the initial/boundary problem for plume dynamics is:

$$\begin{array}{l}\partial_t[s_e(1-z)^{2.5}] + \partial_z[s_e(1-z)^{2.5}u] = 0, \quad \partial_t u + u\partial_z u = \tau \quad (\psi = 1-z) \\ t=0: \; u=0, \; s=0 \; (z>0), \quad z=0: \; u=u_0, \; s=s_0 \; (t \geq 0)\end{array}. \tag{4.10}$$

Initial/boundary conditions in (4.10) indicate that the plume emerged vertically from the spot with initial area $s_0$, is located at the ocean level.

The parameters $s_0$ and $u_0$ will be considered below as constants, although they could generally be time dependent. Thus very shortly, within about couple of minutes after emerging, the plume could be considered as stationary. This stationary plume could exists if it is somehow "feed" from the environment, e.g. by the approaching horizontal warm streak through a boundary layer specific for the plume, in the same manner as described in the previous paper [3] of the series. The stationary solution of (4.10) denoted by asterisks has the form:

$$u_*(z) = u_0\sqrt{1+\kappa z}, \quad s_*(z) = s_0\left((1-z)^{2.5}\sqrt{1+\kappa z}\right)^{-1} \quad (\kappa = 2\tau/u_0^2) \tag{4.11}$$

Additionally, in the adiabatic case temperature is linearly distributed with height: $T(z) = T_0(1-z)$. The solution (4.11) exists within the condensation height of plume $0 < z < z_c$, where $z_c = 1 - T_c/T_0$. If $0 < z_m < z_c$, where $z_m$ is the possible point of minimum jet radius, then the jet cross-section decreases in the interval $0 < z < z_m$ and increases in the interval $z_m < z < z_c$. This situation happens for small values of $u_0$. For large values of $u_0$ the jet radius monotonously decreases.

We now assume that the wind blowing in the area of emerged plume has a horizontal shear $\dot{\gamma}$ (or vertical component of vorticity), which is almost constant from the top to the bottom of the plume. Then at certain time $t_*$ the plume could capture this vorticity and after relatively short restructuring acquires the full constant angular momentum $M_0$ with the following property: $M_0 = [\omega_*(z) + \Omega]s_*(z)$. Here $s_*(z)$ is given in (4.11) and $\Omega$ is the local angular velocity of Earth rotation on $\beta$-plane. We can evaluate the vertical distribution of the angular velocity $\omega_*(z)$ from the approximate equality $M_0 \approx \dot{\gamma}s_0 \approx [\omega_*(0) + \Omega]s_0$. The process of restructuring from plume to initial evolution of hurricane could be described using the non-steady equations (4.7a-c) with following initial and boundary conditions:

$$t = t_*: \ u = u_*(z), \ s = s_*(z) \ (z > 0), \quad z = 0: \ u = u_0(t), \ s = s_0(t) \ (t \geq t_*) \tag{4.12}$$

As discussed before, the restructuring process will be over in few minutes and after that the long process of hurricane maturing will take place.

## 4.4. Hurricane maturing

During this last stage of hurricane evolution, a boundary layer near the ocean interface begins to forming in emerged rotating plume. Also, during this stage non-steady boundary conditions in (4.12) are slowly changing with time. It happens because of slow radial propagation of rotation into the ambient region which was initially at rest, with increasing the external hurricane radius $r_a(t)$. It should be noted that with the radial propagation stage, the airflow in the jet increases because of expansion of region of evaporation at the hurricane bottom in the warm sea.

The quasi-steady state approach used in paper [2] results in the algebraic relation for local angular momentum conservation as an "adiabatic invariant", slowly changing with time:

$$M(r,t) = (\omega + \Omega)s = \chi(t/t_s) \quad (t_r << t << t_s) \tag{4.13}$$

Here $s = r^2$, $\Omega$ is the local angular velocity of Earth rotation on $\beta$ − plane, $t_r$ and $t_s$ are the characteristic time scales for fast and slow motions in the hurricanes, respectively. Equation (4.13) is valid for the whole area $r_0 \leq r \leq r_a$ outside the jet, and possible changes $r_a(t)$ and $r_0(t)$ are considered as extremely slow, i.e. kept them as "frozen" variables. For simplicity, we consider the initial jet radius $r_0$ keeping constant and equal to the external radius of forming EW jet or plume. Thus in the following we will study a cylindrical model for external radius $r_a(t)$ of hurricane with fixed internal radius $r_0$, viewed as the bottom radius of plume.

Equation (4.13) extended to the hurricane boundary $r = r_a(t)$, is rewritten as:

$$M(t) = [\omega(t) + \Omega]r_a^2(t). \tag{4.14}$$

Using (4.14) we can also express the absolute $u_{\varphi a}(t)$ and relative $u_\varphi(t)$ rotational velocities (relative to the Earth rotation) via $M(t)$ as:

$$u_{\varphi a} = M/r_a, \qquad u_\varphi = M/r_a - \Omega r_a. \tag{4.15}$$

The objective of this Section is to describe the slow evolution of external radius $r_a(t)$ and the local angular momentum $M(t)$. We neglect in our modeling the loss of energy spent for the wave generation on the sea surface underlying the hurricane, because the ratio of loss to existing kinetic energy of wind is $\sim h_s/H$ ratio. Here $h_s$ is the height of the hurricane boundary layer and $H$ is the total hurricane height.

We start with introducing two prognostic evolution equations:

$$dr_a/dt \approx ku_\varphi = k(M/r_a - \Omega r_a) \tag{4.16a}$$

$$dM/dt \approx u_{\varphi a}dr_a/dt \quad (k = const \sim 1). \tag{4.16b}$$

The initial conditions for equations (4.16a, b) are:

$$r_a(0) = r_0, \quad M(0) = M_0 = (\omega_0 + \Omega)r_0^2 \quad (\omega_0 \approx \dot{\gamma}_0). \tag{4.17}$$

Here $\dot{\gamma}_0$ is the horizontal shear of wind initiated the plume rotation.

Equation (4.16a) is a phenomenological description of propagation of jump in the relative rotational velocity at the boundary $r = r_a$ due to the Kelvin-Helmholtz instability. This propagation could happen in both inward and outward radial directions. Therefore this problem might be quite different from the common entrainment of liquid motion in the outward direction due to stirring the liquid, when one should consider the modulus of the right-hand side value. An approximate character of this equation is related to its linearity. It seams, however, be good for the slow motions. The value of constant dimensionless parameter $k$ in this equation is unknown and considered below as a fitting parameter of order of unity. Equation (4.16b) assumes the radius growth/decrease due to the radial propagation of unstable boundary as the dominant contribution in the growth/decrease of angular momentum. Only this effect is considered below.

Consider initially the solution of problem (4.16a, b) for radial propagation of a solitary rotational impulse imposed on the plume. Equation (4.16b) can be rewritten as:

$$dM = r_a du_{\varphi a} = (\omega + \Omega) r_a dr_a = (M/r_a) dr_a.$$

Along with the second boundary condition in (4.17), this equation is reduced to the problem:

$$dM/dr_a = M/r_a; \quad M\big|_{r_0} = M_0 = (\omega_0 + \Omega)r_0^2 \tag{4.18}$$

Here $r_0$ is a radius of the plume after rotational restructuring and $\omega_0 = const$ is the angular velocity of the rotational solitary impulse. The sign (or direction of rotation) of angular velocity $\omega_0$ is arbitrary, because it is the transferred to the plume from the arbitrary directed rotational impulse of horizontally sheared wind. Recall that the positive (cyclonic or anti-clock wise) value of $\omega_0$ has the same sign as $\Omega$, and negative (anti-cyclonic) value of $\omega_0$ acquires the clock wise rotation in Northern hemisphere. Solution of (4.18) is:

$$M(t) = M_0 r_a(t)/r_0 = (\omega_0 + \Omega) r_0 r_a(t). \tag{4.19}$$

Substituting (4.19) into (4.16a) yields the problem:

$$dr_a/dt + k\Omega r_a = (\omega_0 + \Omega)r_0, \quad r_a(0) = r_0. \tag{4.20}$$

Then the solution of (4.19)-(4.20) is:

$$r_a(t) = r_0[1 + (\omega_0/\Omega)(1 - e^{-\Omega k t})] \tag{4.21a}$$

$$M(t) = (\omega_0 + \Omega)r_0^2[1 + (\omega_0/\Omega)(1 - e^{-\Omega k t})] \ . \tag{4.21b}$$

As follows from (4.21a,b) $u_{\varphi a} = M(t)/r_a(t) = (\omega_0 + \Omega)r_0 = const$, i.e. equation (4.16b) holds precisely. It means that in the non-rotating frame of reference the tangential velocity of rotation is constant at the jump, exactly as in the plane case of Kelvin-Helmholtz instability.

Formulas (4.21a,b) allow easy analysis of maturing stage of hurricane depending on sign $\omega_0$.

(i) If $\omega_0 > 0$, both $r_a(t)$ and $M(t)$ monotonically increase to their stationary values $r_s = r_a(\infty)$ and $M_s = M(\infty)$, where

$$r_s = r_0(1 + \omega_0/\Omega), \quad M_s = \Omega r_0^2(1 + \omega_0/\Omega)^2 = \Omega r_s^2. \tag{4.22}$$

Formulae (4.22) show that at the radius $r_s$ the relative angular rotation $\omega_s = 0$. They also clearly demonstrate that maximal (stationary) value $u_{\varphi m}$ of hurricane tangential wind speed, achieved at $r = r_0$ when $t \to \infty$, increases with the increase in value of $\omega_0$ as:

$$u_{\varphi m} = M_s/r_0 - \Omega r_0 = \omega_0 r_0(2 + \omega_0/\Omega). \tag{4.23}$$

(ii) If $\omega_0 < 0$, the disturbances propagate inwards and the evolution of radius of rigid rotation is described by evolution of $r_a(t)$, i.e. in this case $r_0(t) = r_a(t)$. Then both $r_a(t)$ and $M(t)$ monotonically decrease to their stationary values, $r_s = r_a(\infty)$, $M_s = M(\infty)$. Here

$$r_s = r_0 \begin{cases} 1 - |\omega_0|/\Omega, & (|\omega_0| < \Omega) \\ 0, & (|\omega_0| \geq \Omega) \end{cases}, \quad M_s = \begin{cases} \Omega r_s^2, & (|\omega_0| < \Omega) \\ 0, & (|\omega_0| \geq \Omega) \end{cases} \tag{4.24}$$

It is seen that in the first case in (4.24), the relative angular velocity decreases to the steady value $\omega_{s0} = 0$ for "infinite" time. In the second case in (4.24) when $|\omega_0| \geq \Omega$, the collapse of initial plume happens for a finite time, whose value is elementary found from formulas (4.21).

The above analysis shows that the model selects for developing tropical cyclone (TC) only those rotating jets that have cyclonic rotations imposed on the initial plumes, while the jets with anti-cyclonic rotations initiated by horizontally sheared wind will eventually die.

A brief comparison of relations (4.13) and (4.21) shows that the characteristic "slow" time of evolution of hurricane $t_s \sim 1/\Omega$. It means that the proposed model can be considered as consistent with the quasi-steady-state approximation.

Only the evolution of a rotating jet under a solitary action of horizontally sheared wind fluctuation has been analyzed above. Nevertheless, in real situation, there might be multiple fluctuations randomly applied to the same previously initiated TC. Then the TC evolution with interference of multiple rotations can be easily described by the proposed model, even including their random character, bearing in mind that the angular velocities of rigid rotation in the TC could be increased or decreased by the interactive wind fluctuations. In this case the initiated TC will show a slow evolution during an induction time and waiting for a large wind fluctuation which could produce a large enough value of $\omega_0$ Similar observations have been described in [4], Ch.2.42. However, those large values could happen relatively seldom. This perhaps is the reason why the hurricanes emerged not that often.

According to Ref.[4] (Ch.2.4.2), the TC can be transformed into a hurricane during 5-6 days after the action of wind with vorticity $\omega_0 \approx 10^{-4} s^{-1}$.

We now make model estimations using the following values of parameters:

$$k \approx 1,\ \Omega = 5.5 \times 10^{-6} s^{-1},\ \omega_0 = 10^{-4} s^{-1},\ r_0 = 30 km.$$

With these data, the model calculations yield:

(i) The characteristic time of hurricane development: $t_s \approx 3/\Omega \approx 5.45 \times 10^5 \sec \approx 6.3$ days ;

(ii) Characteristic radius of developed hurricane: $r_s = r_0(1 + \omega_0/\Omega) \approx 575 km$ ;

(iii) Maximum speed of developed hurricane: $u_{\varphi \max} = \omega_0 r_0 (2 + \omega_0/\Omega) \approx 60.5 m/s$ .

(iv) Grow of angular momentum: from $M_0 \approx 9.49 \times 10^4$ to $M_s \approx 1.82 \times 10^6\ m^2/s$ .

These results seem to be quite realistic.

## 4.5. Remarks on tornados

Many parts of developed theoretical models are also applicable to tornados. Tornados can emerge on land and on sea surface. Unlike hurricanes tornados have relatively small jet radius but so high intensity of angular rotation $\omega_0^T <~ 1\sec^{-1}$ that it is practically independent of local angular rotation of Earth. This is the first remarkable difference between hurricanes and tornados.

Highly probable that existing horizontally sheared wind with local vertical vortices of a high intensity cause these high angular velocities. Non-steady equations (4.7a-c), derived for initial stage of hurricane evolution, seem to be applicable to evolution of tornado too. But unlike very long time needed for hurricane maturing, one can expect using (4.9) very quick (about several minutes long) period of tornado maturing, following by relatively long (say, several hours - one day) quasi-stationary horizontal travel. This is the second remarkable difference between hurricanes and tornado. It seems that the reason for that is a much higher friction of tornados even at sea.

The steady structure of upper part of tornado seems to be described well by the stationary version of equations (4.7a-c). Since tornados exist relatively short time, the eye region inside of EW jet could be absent there, although a little is known of the internal tornado structure. A little is known either about the tornado boundary layer and the balances of heat, mass and momentums there. The features of sea type of tornado are more similar to the hurricanes traveling over open sea, but because they are less harmful, their studies seem to be of small interest. On the other hand, land type tornados are so sporadic and intense, that their experimental studies are not only difficult but also very dangerous. If tornados travel like hurricanes towards the warm air bands, the sporadic behavior of the land tornados could be explained by irregularities in the locations of these bands.

In case of quasi-steady tornado, one can use formula (4.4) for radial distribution of angular momentum $M_T$ with "infinite" limit for its external radius ($r_a^T \to \infty$). Because of that the quasi-steady tangential velocity $u_\varphi^T$ decays $\sim 1/r$. The radial distribution of tangential velocity $u_\varphi(r)$ is presented due to (4.4) as:

$$u_\varphi(r,z) \approx u_{\varphi m} \times \begin{cases} r/r_e, & 0 < r \leq r_e \\ r_e/r, & r_e \leq r \leq r_a \end{cases}, \quad u_{\varphi m} = \frac{M_e}{r_e}. \qquad (4.25)$$

Respective surface pressure distribution $p_s(r)$ is presented due to (4.5) as:

$$p_s / p_a^0 = 1 - \frac{u_{\varphi m}^2}{2RT_a^0}\left[1 + \frac{T_0}{T_a^0}\left(1 - \frac{r^2}{r_e^2}\right)\right] \qquad (0 \leq r \leq r_e)$$

$$p_s / p_a^0 = 1 - \frac{u_{\varphi m}^2}{2RT_a^0}\left(\frac{r_e}{r}\right)^2 \qquad (r \geq r_e) \qquad (4.26)$$

In (4.26) we used the relation $\rho \approx p_a/(RT_a)$, and upper zero index indicates the surface values of ambient pressure and temperature. The surface pressure $p_{s0} = p_s(0)$ at the tornado's center characterizing the "tornado depression" is expressed due to (4.26) as:

$$p_{s0} / p_a^0 = 1 - \frac{u_{\varphi m}^2}{2RT_a^0}\left(1 + \frac{T_0}{T_a^0}\right). \qquad (4.27)$$

The following example illustrates formulas (4.25) – (4.27), using extreme value of angular velocity $\omega_0^T \approx 1 \sec^{-1}$ for tornados and a typical value for bottom jet radius $r_e^T \approx 70m$. The value of angular momentum is calculated as $M_T \approx 5 \times 10^3 m^2/\sec$. Then the values of tangential velocity at the external radius $r_e^T$ of tornado jet, and at the distance $r_a^T \approx 1000m$ are calculated as: $u_{\varphi m} = \omega_0^T r_e^T = 70 m/\sec = 252 km/hour = 160 mph$, and $u_\varphi(r_a^T) = \omega_0^T (r_e^T)^2 / r_a^T \approx 5$ m/sec $= 18\ km/hour = 11 mph$.

The tornado depression can be evaluated using (4.27). When neglecting the temperature differences, i.e. approximate $T_0/T_a^0 \approx 1$, formula (4.27) takes the form:

$$p_{s0} \approx p_{sa}\left(1 - \frac{u_\varphi^2(r_e^T)}{RT_a^0}\right). \qquad (4.28)$$

For above value $u_\varphi(r_e^T) = 70 m/\sec$ and standard values of thermodynamic parameters, formula (4.28) gives the value of tornado depression: $p_{s0} \approx 944 mb$. The vacuum related to this depression is so high that at the circular area $A = \pi r_0^2$ at the tornado surface with the radius $r_0 = 1m$ creates the lifting force $F = A(p_{sa} - p_{s0}) \approx 1.77 \, tons$.

## V.6. Conclusions

This paper mostly developed a model of hurricane geneses based on the three-step scenario: (i) formation of plume, (ii) converting the plume to rotating jet by plume's acquiring the rotation from the wind with horizontal shear.

Based on simplified average approach to rotating jet, modeling non-steady equations suitable for future numerical studies were derived for analyzing initial stage of hurricane. Scaling of these equations revealed a fast restructuring in emerged hurricane taking about several minutes.

A simple analytical model, based on radial propagation of Kelvin-Helmholtz instability, has been developed for long time hurricane maturing. It was shown that the model demonstrates very realistic behavior. Especially important is that radial expansion of emerged hurricane happens only for cyclonically directed initial rotating disturbance, acquired from the horizontally sheared wind.

Many aspects of the models developed for hurricanes are also suitable for tornados. Some features of tornados have been discussed in the previous Section V.5, where an example of extreme quasi-steady tornado illustrates remarkable effects.

## References

1. A.I. Leonov, *Aerodynamic models of hurricanes*. I. Model description and principles of horizontal motions of hurricane (the first paper of the Series).
2. A.I. Leonov, *Aerodynamic models of hurricanes*. II. Model of upper hurricane layer (the second paper of the Series).
3. A.I. Leonov, *Aerodynamic models of hurricanes*. IV. Modeling hurricane boundary layer (the third paper of the series)
4. R. A. Anthes, *Tropical Cyclones, Their Evolution, Structure and Effects*, Am. Met. Soc., Science Press, Ephrata, PA (1982).
5. K.V. Ooyama, Numerical simulations of life cycle of tropical cyclones, *J. Atmos. Sci.*, **26**, 3-40 (1983).
6. K.V. Ooyama, Conceptual evolution of the theory and modeling of the tropical cyclone. *J. Met. Soc. Japan*, **60**, 369-379 (1982).

7. K. A. Emanuel, The finite amplitude nature of tropical cyclogenesis, *J. Atmos. Sci.*, **46**, 3431-3456 (1989).
8. K. A. Emanuel, Some aspects of hurricane inner-core dynmics and energetics, *J. Atmos. Sci.*, **54**, 1014-1026 (1997).
9. M.T. Montgomery, V.A. Vladimirov and P.V. Denisenko, An experimental study of hurricane mesovortices, *J. Fluid. Mech.*, **471**, 1-32 (2002).